\documentclass[fleqn,multphys,vecphys]{svmult}

\usepackage{makeidx}   
\usepackage{graphicx}  
\makeindex             
 \usepackage{hyperref}  

\newcommand{\iDev}[1]{#1}
\newcommand{\iName}[1]{#1}
\newcommand{\iStreet}[1]{#1}
\newcommand{\iPostcode}[1]{#1}
\newcommand{\iCity}[1]{#1}
\newcommand{\iCountry}[1]{#1}
\newcommand{\email}[1]{\texttt{#1}}

\newcommand{\bra}[1]{\left\langle #1\right|}
\newcommand{\ket}[1]{\left|#1\right\rangle }

\def\abstractname{Abstract.}%

\begin{document}

\title*{Quantum-Dot Spin Qubit and Hyperfine Interaction}          
\toctitle{Quantum-Dot Spin Qubit and Hyperfine Interaction}        
\titlerunning{Quantum-Dot Spin Qubit and Hyperfine Interaction}    

\author{D. Klauser, W. A. Coish and Daniel Loss}                   

\authorrunning{D. Klauser et al.} 

\institute{\iDev{Department of Physics and Astronomy},                  
\iName{University of Basel}, 
\iStreet{Klingelbergstrasse 82},\newline
\iPostcode{CH-4056}
\iCity{Basel},
\iCountry{Switzerland}\newline
\email{}
}

\maketitle

\begin{abstract}
We review our investigation of the spin dynamics for two electrons
confined to a double quantum dot under 
the influence of the hyperfine interaction between the electron spins
and the surrounding nuclei. Further we propose a scheme to narrow the
distribution of difference in polarization between the two dots in
order to suppress hyperfine induced decoherence.
\end{abstract}

\section{Introduction\label{sec:introduction}}
\footnotetext[1]{Presented as plenary talk at the annual DPG meeting 2006, Dresden
(to appear in Advances in Solid State Physics vol. 46, 2006).}

The fields of semiconductor physics and electronics have been successfully
combined for many years. The invention of the transistor meant a revolution for
electronics and has led to significant development of semiconductor physics and
its industry. More recently, the use of the spin degree of freedom of electrons,
as well as the charge, has attracted great
interest~\cite{awschalom:2002a}. In addition to applications for
spin electronics (spintronics) in conventional devices, for instance based on
the giant magneto-resistance effect~\cite{baibich:1988a} and spin-polarized
field-effect transistors~\cite{datta:1990a}, there are applications that exploit
the quantum coherence of the spin. 
Since the electron spin is a two-level system, it is a natural candidate for the realization of a
quantum bit (qubit)~\cite{Loss:1998a}. A qubit is the basic unit of information
in quantum computation. The confinement of electrons in semiconductor structures like
quantum dots allows for better control and isolation of the electron spin from
its environment. Control and isolation are important issues to consider for the design
of a quantum computer.
 
Formally, a quantum computation is performed through a set of transformations,
called gates~\cite{preskill}. A gate applies a unitary transformation $U$
to a set of qubits in a quantum state $|\Psi\rangle$. At the end of the
calculation, a measurement is performed on the qubits (which are in the state
$|\Psi'\rangle=U\,|\Psi\rangle$). There are many ways to choose sets
of universal quantum gates. These are sets of gates from which any
computation can 
be constructed, or at least approximated as precisely as desired. Such a set
allows one to perform any arbitrary calculation without inventing a new gate
each time. The implementation of a set of universal gates is therefore of
crucial importance. It can be shown that it is possible to construct such a set
with gates that act only on one or two qubits at a time~\cite{Barenco:1995a}.

The successful implementation of a quantum computer demands that some basic
requirements be fulfilled. These are known as the DiVincenzo
criteria~\cite{divincenzo:2000a} and can be summarized in the following:

\begin{enumerate}
\item{{\em Information storage\---the qubit:} We need to find some quantum
property of a scalable physical system in which to encode our bit of
information, that lives long enough to enable us to perform computations.}
\item{{\em Initial state preparation:} It should be possible to set the state of
the qubits to 0 before each new computation.}
\item{{\em Isolation:} The quantum nature of the qubits should be tenable; this
will require enough isolation of the qubit from the environment to reduce the
effects of decoherence.}
\item{{\em Gate implementation:} We need to be able to manipulate the states of
individual qubits with reasonable precision, as well as to induce interactions
between them in a controlled way, so that the implementation of gates is
possible. Also, the gate operation time $\tau_s$ has to be much shorter than the
decoherence time $T_2$, so that $\tau_s/T_2\ll r$, where $r$ is the maximum tolerable
error rate for quantum error correction schemes to be effective.}
\item{{\em Readout:} It must be possible to measure the final state of our
qubits once the computation is finished, to obtain the output of the
computation.}
\end{enumerate}

To construct quantum computers of practical use, we emphasize that the
scalability of the device should not be overlooked. This means it should be
possible to enlarge the device to contain many qubits, while still adhering to
all requirements described above. It should be mentioned here that this
represents a challenging issue in most of the physical setups proposed so
far.

\section{Quantum-Dot Spin Qubit\label{sec:spinqubit}}

The requirement for scalability motivated the Loss-DiVincenzo proposal
\cite{Loss:1998a} for a solid-state quantum computer based on electron spin
qubits. 
\begin{figure}
\begin{center}
\scalebox{0.2}{\includegraphics{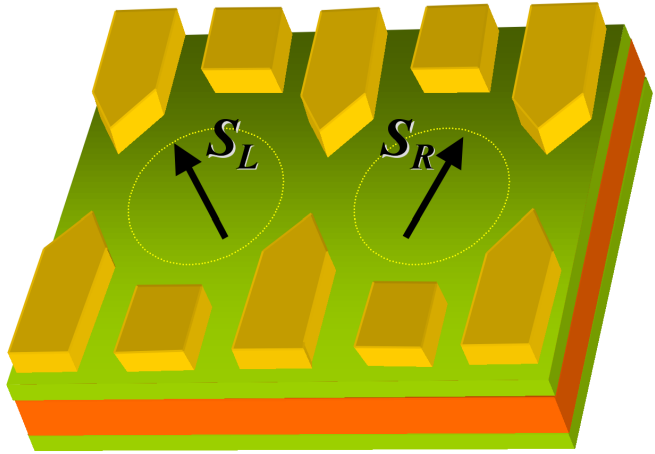}}
\end{center}
\caption{\label{fig:twoqubitexchange} Two neighbouring electron spins confined
  to quantum dots, as in the Loss-DiVincenzo proposal.  The lateral confinement
  is controlled by top gates.  A time-dependent Heisenberg exchange 
  coupling $J(t)$ can be pulsed high by pushing the electron spins closer, 
  generating an appreciable overlap between the neighbouring orbital wave
  functions.} 
\end{figure}

The qubits of the Loss-DiVincenzo quantum computer are formed from the two spin
states ($\left|\uparrow\right>,\left|\downarrow\right>$) of a confined electron.
The considerations discussed in this proposal are generally applicable to
electrons confined to any structure, such as atoms, molecules, etc., although
the original proposal focuses on electrons localized in quantum dots.  These
dots are typically generated from a two-dimensional electron gas (2DEG), in
which the electrons are strongly confined in the vertical direction.  Lateral
confinement is provided by electrostatic top gates, which push the electrons
into small localized regions of the 2DEG (see Fig. 
\ref{fig:twoqubitexchange}). Initialization of the quantum computer
can be achieved 
by allowing all spins to reach their thermodynamic ground state at low
temperature $T$ in an applied magnetic field $B$ (i.e., virtually all spins will
be aligned if the condition $\left|g\mu_\mathrm{B}B\right| \gg k_\mathrm{B}T$ is
satisfied, with $g$-factor $g$, Bohr magneton $\mu_\mathrm{B}$, and Boltzmann's
constant $k_\mathrm{B}$).  Single-qubit
operations can be performed, in principle, by changing the local effective
Zeeman interaction at each dot individually.  To do this may require large
magnetic field gradients \cite{wu:2004a}, $g$-factor engineering
\cite{medeiros-ribeiro:2003a}, magnetic layers, the inclusion of
nearby ferromagnetic dots \cite{Loss:1998a}, polarized nuclear
spins, or optical schemes.

In the Loss-DiVincenzo proposal, two-qubit operations are performed by pulsing the electrostatic
barrier between neighboring spins.  When the barrier is high, the spins are
decoupled.  When the inter-dot barrier is pulsed low, an appreciable overlap develops
between the two electron wave functions, resulting in a non-zero Heisenberg
exchange coupling $J$.  The Hamiltonian describing this time-dependent process
is given by 
\begin{equation} 
\label{eqnExchangeHamiltonian}
H(t) = J(t)\mathbf{S}_L\cdot\mathbf{S}_R.
\end{equation} 
This Hamiltonian induces a unitary evolution given by the
operator $U=\mathcal{T}\exp\left\{-i\int H(t) dt/\hbar \right\}$, where
$\mathcal{T}$ is the time-ordering operator.  If the exchange is pulsed on for a
time $\tau_\mathrm{s}$ such that $\int J(t) dt/\hbar=J_0\tau_s/\hbar=\pi$, the states of
the two spins, with associated operators $\mathbf{S}_L$ and $\mathbf{S}_R$, as
shown in figure \ref{fig:twoqubitexchange}, will be exchanged.  This is the {\sc
swap} operation.  Pulsing the exchange for the shorter time $\tau_\mathrm{s}/2$ generates
the ``square-root of {\sc swap}'' operation, which can be used in conjunction
with single-qubit operations to generate the controlled-{\sc not} (quantum {\sc
xor}) gate \cite{Loss:1998a}.  
The ``square-root of {\sc swap}'' gate has recently been implemented
in an experiment by Petta \emph{et al.} \cite{Petta:2005b} with a switching 
time $\tau_s=180$ ps. 
For scalability, and application of quantum error correction
procedures in any quantum computing proposal, it is important to turn off
inter-qubit interactions in the idle state.  In the Loss-DiVincenzo proposal,
this is achieved with exponential accuracy since the overlap of neighboring
electron wave functions is exponentially suppressed with increasing separation. A
detailed investigation of decoherence during gating due to a bosonic environment
was performed in the original work of Loss and DiVincenzo.  Since then, there
have been many studies of leakage and decoherence in the context of the
quantum-dot quantum computing proposal. 

In addition to the interaction-based gate operations introduced above, it has
been shown recently \cite{Engel:2005a,beenakker:2004a} that it is also possible to
generate the controlled-{\sc not} based on partial Bell state
(parity) measurements.  

For both interaction-based and measurement-based quantum computation
with the quantum-dot spin qubit,
decoherence due to the coupling of the qubit to its environment is a major obstacle. There
are two important sources of decoherence in GaAs quantum dots: spin-orbit coupling
(interaction between spin and charge fluctuations) and hyperfine
coupling (interaction between the electron spin and nuclear
spins). In the case of spin-orbit interaction alone it has been shown that the
decoherence time $T_2$ (which is the relevant timescale for quantum
computing tasks) exceeds the relaxation time $T_1$ and is given by
$T_2=2 T_1$ to leading order in the spin-orbit coupling \cite{golovach:2004a}. 
Since the $T_1$ obtained in measurements
\cite{elzerman:2004a,Kroutvar:2004a} is on the order of ms, but the
ensemble-averaged dephasing time 
$T_2^*$ measured is $\sim 10$ ns, spin-orbit interaction is not
limiting for the dephasing time $T_2^*$. The limiting source of
decoherence is the hyperfine interaction \cite{Petta:2005b}. 

\section{Hyperfine Interaction in Single and Double Dots}

The hyperfine interaction between the electron spin and the nuclear spins present in all 
III-V semiconductors \cite{Schliemann:2003a} leads to the strongest 
decoherence effect 
\cite{Petta:2005b,Burkard:1999a,Erlingsson:2001a,Erlingsson:2002a,Khaetskii:2002a,Merkulov:2002a,
Khaetskii:2003a,Coish:2004a,Coish:2005a}. 
Experiments \cite{Petta:2005b,Bracker:2005a,Dutt:2005a,Koppens:2005a} have yielded values 
for the free-induction spin dephasing time $T_2^*$ that are
consistent with $T_2^*\sim \sqrt{N}/A \sim  
10 \mathrm{ns}$ \cite{Khaetskii:2002a,Merkulov:2002a,Khaetskii:2003a}
for $N = 10^6$, $\hbar=1$, and 
$A=90\mu \mathrm{eV}$ in GaAs, where $N$ is the number of nuclei within one quantum dot
Bohr radius and $A$ characterizes the weighted average hyperfine
coupling strength in GaAs \cite{Paget:1977a}. 
This is to be contrasted with potential spin-echo envelope decay times, which may be much longer 
\cite{Coish:2004a,Sousa:2003a,Shenvi:2005a,Yao:2005a}. With a two-qubit switching time of 
$\tau_s\sim 180$ ps \cite{Petta:2005b} this only allows $\sim 10^2$
gate operations within $T_2^*$, which
falls short (by a factor of $10$ to $10^2$) of current requirements for efficient quantum 
error correction \cite{Steane:2003a}.

There are several ways to overcome the problem of hyperfine-induced decoherence, of 
which measurement and thus projection of the nuclear spin state may be the most
promising \cite{Coish:2004a}. Other methods include polarization
\cite{Burkard:1999a,Khaetskii:2003a,Coish:2004a,Imamoglu:2003a} of the nuclear spins and
spin echo techniques \cite{Petta:2005b,Coish:2004a,Shenvi:2005a}. However, in order to 
extend the decay time  by an order of magnitude through polarization 
of the nuclear spins, a polarization of  above 99\% is required \cite{Coish:2004a}, but the best 
result so far reached is only $\sim$60\%  in quantum dots \cite{Bracker:2005a}. 
With spin-echo techniques, gate operations still must be performed within the single-spin 
free-induction decay time, which requires faster gate operations. A projective
measurement of the nuclear spin state leads to an extension of the
free-induction decay time for the spin. This extension is only
limited by the ability to do a strong measurement since the longitudinal nuclear spin in a 
quantum dot is expected to 
survive up to the spin diffusion time, which is on the order of seconds for nuclear spins
surrounding donors in GaAs \cite{Paget:1982a}.

A detailed analysis of the spin dynamics for one electron in a single
quantum dot can be found in Ref.\cite{Coish:2004a}. Here we
concentrate on the case of two electrons in a double quantum dot. 
The spin $\mathbf{S}_l$ of an electron in quantum dot $l=1,2$,
interacts with the surrounding nuclear spins $\mathbf{I}_k$  
via the Fermi contact hyperfine interaction:
\begin{equation}\label{Eq:fermihyperfine}
H_{\mathrm{hf}}=\mathbf{S}_l\cdot \mathbf{h}_l;\,\,\,
\mathbf{h}_l=\sum_k A^l_k \mathbf{I}_k;\,\,\,
A^l_k=v_0A\left|\psi^l(\mathbf{r}_k)\right|^2, 
\end{equation}
where $v_0$ is the volume of a crystal unit cell containing one
nuclear spin. 
The effective Hamiltonian in the subspace
of one electron on each dot is best written in terms of the sum and
difference of electron spin and collective nuclear spin operators:
$\mathbf{S}=\mathbf{S}_1+\mathbf{S}_2,\delta\mathbf{S}=\mathbf{S}_1-\mathbf{S}_2$
and $\mathbf{h}=\frac{1}{2}(\mathbf{h}_1+\mathbf{h}_2),
\delta\mathbf{h}=\frac{1}{2}(\mathbf{h}_1-\mathbf{h}_2)$:
\begin{equation}
H_{\mathrm{eff}}(t)=\epsilon_z S^z +\mathbf{h\cdot S}+
\delta\mathbf{h\cdot}\delta\mathbf{S} +\frac{J(t)}{2}\mathbf{S\cdot S}-J(t),
\end{equation}
where $\epsilon_z=g\mu_B B$ is the Zeeman splitting induced
by an applied magnetic field $\mathbf{B}=(0,0,B), B>0$. 
We assume that the Zeeman splitting is much larger than $\langle
\delta \mathbf{h}\rangle_{\mathrm{rms}}$ and $\langle\mathbf{h}_i 
\rangle_{\mathrm{rms}}$, where $\langle\mathcal{O}\rangle_{\mathrm{rms}}
=\bra{\psi_I} \mathcal{O}^2\ket{\psi_I}^{1/2}$ is the
root-mean-square expectation value of the operator $\mathcal{O}$
with respect to the nuclear spin state $\ket{\psi_I}$. Under these
conditions the relevant spin Hamiltonian becomes block diagonal  with
blocks labeled by the total electron spin projection along the
magnetic field $S^z$. In the subspace of $S^z=0$ the Hamiltonian can
be written as ($\hbar=1$) \cite{Coish:2005a,Klauser:2006a}  
\begin{equation}\label{Hamiltonian}
H_{0}(t)=\frac{J(t)}{2}\tau^{x}-\frac{1}{2}\Omega
\tau^z;\,\,\,J(t)=J_0+j\cos(\omega t),\,\,\,\Omega=2(\delta h^z+\delta b^z).
\end{equation}
Here, $\delta b^z$ is the inhomogeneity of an externally
applied classical static magnetic field with $\delta b^z \ll B$. The
Pauli matrices $\mathbf{\tau}^\alpha,\,\alpha=x,y,z$ are given in the basis of
$\ket{+}\equiv\ket{\tau^z=1}=\ket{\downarrow\uparrow}$ and
$\ket{-}\equiv\ket{\tau^z=-1}=\ket{\uparrow\downarrow}$.

The dynamics of the two-electron spin states depends strongly on the
initial state of the nuclear spin system. We denote by $\ket{n}$ the
eigenstates of $\delta h^z$ with  $\delta h^z \ket{n}=\delta h^z_n
\ket{n}$. If the initial state of the nuclear spin 
system is $\rho_I(0)=\ket{n}\bra{n}$ and if we neglect spin-flip
processes (as can be done for a large enough magnetic field $B$), then
the initial spin state of the electron does not decay. Thus, if it is
possible to prepare the nuclear spin system in an eigenstate $\ket{n}$,
hyperfine-induced decoherence could be overcome. In general, however,
the initial state of the nuclear spin system is not an eigenstate
$\ket{n}$ but a general mixture:
\begin{equation}
\rho_I(0)= \sum_ip_i\ket{\psi_I^i}\bra{\psi_I^i};\,\,\,\ket{\psi_{I}^i}=\sum_na_n^i\ket{n},
\end{equation}
where the $a_n^i$ satisfy the normalization condition
$\sum_n|a_n^i|^2=1$ and $\sum_i p_i=1$. We denote by
$\rho_I(n)=\sum_{i}p_i|a_n^i|^2$ the diagonal  
elements of the nuclear spin density operator. 

For a large number of nuclear spins $N\gg1$ which are in a superposition of
$\delta h^{z}$-eigenstates $\ket{n}$,  $\rho_{I}(n)$
describes a continuous Gaussian distribution of $\delta h_{n}^{z}$
values, with mean $\overline{\delta h^{z}}$ and variance $\sigma^{2}=
\overline{\left(\delta h^{z}-
\overline{\delta h^{z}}\right)^{2}}$.
In the limit of large $N$, the approach to a Gaussian distribution 
for a sufficiently randomized nuclear system is guaranteed by the
central limit theorem \cite{Coish:2004a}. We perform the continuum limit according to
\begin{eqnarray}
\sum_{n}\rho_{I}(n)f(n)  \to  \int dx
\rho_{I}(x)f(x);\label{Eq:continuumlimit}\,\,\,
\rho_{I}(x)  =  \frac{1}{\sqrt{2\pi}\sigma}\label{Eq:gaussian}
\exp\left(-\frac{\left(x-x_0\right)^{2}}
{2\sigma_0^{2}}\right),\end{eqnarray}
where $x=\delta h_n^z+\delta b^z$, $x_0=\overline{\delta
  h^z}+\delta b^z$ and $\sigma_0^2=\overline{x^2} 
-x_0^2$.

For the case of a static exchange interaction $J(t)=J_0$, the decay of
the two-electron spin states in the $S^z=0$ subspace due to the
Gaussian distribution of nuclear spin states may be calculated in
several interesting limits \cite{Coish:2005a,Klauser:2006a}. Assuming the initial state of the
two-electron system is $\rho_e(0)=\ket{+}\bra{+}$, the probability
$P^+$ to measure the $\ket{+}$ state as a function of time is given by 
\begin{equation}
P^+_{J=J_0}(t)=\int_{-\infty}^{\infty}\rho_{I}(x) 
\left(\frac{1}{2}+\frac{2x^2}{s(x)^2}+\frac{J^2}{2s(x)^2}\cos(s(x)t)\right)
\end{equation}
with $s(x)=\sqrt{J^2+4x^2}$. In the limit $\sigma_0 \rightarrow 0$, which corresponds 
to one fixed eigenvalue, there is no decay. However, for $\sigma_0 > 0$ there is decay.
For the regime $|x_0|\gg \sigma_0$ we have a Gaussian decay at short times with a 
decay time $t_0 \sim 1/\sigma_0$:
\begin{eqnarray}
P^+_{J=J_0}(t) \approx \frac{1}{2}+\frac{2x_0^2}{\omega_0^2}+
\left(\frac{1}{2}-\frac{2x_0^2}{\omega_0^2}\right)
\exp\left(-\frac{t^2}{2t_0^2}\right)\cos(\omega_0 t),\\
\omega_0=\sqrt{J^2+4x_0^2},\,\,\,t_0=\frac{\omega_0}{4
  |x_0|\sigma_0};\,\,\,|x_0|\gg \sigma_0,\,\,\,t\ll\frac{\omega_0^{3/2}}{2J^2\sigma_0^2}.
\end{eqnarray}
Thus, decreasing $\sigma_0$ increases the
coherence time $t_0$. Hence, the strategy to suppress hyperfine-induced
decoherence is to narrow the Gaussian distribution of nuclear spin
eigenvalues through a measurement of the eigenvalue of $\delta h^z$, i.e., of the difference 
in polarization between the two dots \cite{Coish:2004a,Coish:2005a}.
It has also been proposed to measure the nuclear spin polarization using a phase estimation
method \cite{Giedke:2005a}. In the ideal case, phase estimation yields one bit of information
about the nuclear spin system for each perfectly measured electron. Optical methods have
also been proposed \cite{Stepanenko:2005a}. The all-electrical method 
we propose here can be applied with current technology used in
Refs. \cite{Petta:2005b,Koppens:2005a}.  


\section{Nuclear Spin State Narrowing\label{sec:statenarrowing}}

The general idea behind state narrowing is that the evolution of the two-electron system is 
dependent on the collective nuclear spin state and thus knowing the
evolution of the two-electron  system determines the nuclear spin state. 

The eigenstates of the Hamiltonian $H_0$ are product states: 
if the nuclear spin system is in an eigenstate $\ket{n}$ of $\delta h^z$ with 
$\delta h^z \ket{n}=\delta h^z_n \ket{n}$,
we have $H \ket{\psi}= H_n \ket{\psi_e}\otimes \ket{n}$, where in
$H_n$ the operator $\delta h^z$ has been replaced by $\delta h^z_n$
and $\ket{\psi_e}$ is the electron spin part of the wave
function. Thus, in the Hamiltonian for the evolution of the initial two-electron system,  
the parameter $\delta h^z_n$ is determined by the state of the nuclear
spin system. 
Initializing the two-electron system to the $\ket{+}$ state, i.e.,
$\rho_e(0)=\ket{+}\bra{+}$  and performing a measurement in the $\ket{\pm}$ basis at
time $t_m$ yields for the distribution of nuclear spin eigenvalues
(which is the diagonal part of the nuclear spin density operator in
the continuum limit) after the measurement \cite{Klauser:2006a} 
\begin{eqnarray}
\rho_{I}^{(1,+,\omega)}(x) &=&
\rho_I(x)(1-L_{\omega}(x))\frac{1}{P^{+}_{\omega}}, \label{Eq:rhoplus}\\
\rho_{I}^{(1,-,\omega)}(x) &=&
\rho_I(x)L_{\omega}(x)\frac{1}{P^{-}_{\omega}}, \label{Eq:rhominus}
\end{eqnarray}
where $\rho_I(x)$ is the initial Gaussian distribution of nuclear spin eigenvalues (see Eq. (\ref{Eq:gaussian}))
and the probabilities $P^{\pm}$ for measuring $\ket{\pm}$ are given by
\begin{eqnarray}
P^{+}_{\omega}  &=& \int_{-\infty}^{\infty} dx
\rho_I(x)(1-L_{\omega}(x)), \\
P^{-}_{\omega}  &=& \int_{-\infty}^{\infty} dx
\rho_I(x)L_{\omega}(x),\label{Eq:pminus}
\end{eqnarray}
with
\begin{equation}\label{Eq:lorentzian}
L_{\omega}(x)=\frac{1}{2}\frac{(j/4)^2}{(x-\frac{\omega}{2})^2+(j/4)^2}.
\end{equation}
To obtain this result we have assumed that the measurement is
performed with low time resolution \cite{assumption} $\Delta t \gg
1/j$ and that the parameters satisfy the requirements given in
Eq. (\ref{eq:requirements}) below. The distribution of nuclear spin
eigenvalues after  the 
measurement depends on the result of the measurement (whether
$\ket{+}$ or $\ket{-}$ was measured) and on the driving frequency
$\omega$ of the oscillating exchange interaction $J(t)$. In the case
where the measurement outcome is $\ket{+}$, the initial distribution
$\rho_I(x)$ is multiplied by $1-L_{\omega}(x)$  which causes a dip in
$\rho_I(x)$ at $x=\omega/2$. However, in the case where the result of
the measurement was $\ket{-}$, the initial distribution $\rho_I(x)$ is
multiplied by $L(x)$. The full-width at half-maximum (FWHM) of
$L_{\omega}(x)$ is $j/2$, i.e., half the amplitude of the applied
oscillation exchange interaction $J(t)$. Thus, choosing
$j<\sigma_0$, $\rho_I^{(1,-,\omega)}(x)$ is dominated by the Lorentzian and
therefore the FWHM of the initial nuclear spin distribution has been
narrowed by a factor $\approx j/4\sigma_0$. The probability
$P^-_{\omega}$ to measure $\ket{-}$ in the regime $j\ll \sigma_0$ is
given by $P^-_{\omega}\approx(j/6\sigma_0)\exp((x_0-\omega/2)^2/
2\sigma_0^2)$ and the nuclear spin distribution after measuring
$\ket{-}$ is centered around $\omega/2$. Thus, through such a
measurement it is possible to choose the center of the nuclear spin
distribution by choosing the driving frequency. However, the larger
the difference $x_0-\omega/2$, the smaller is the probability to have
measurement outcome $\ket{-}$, which leads to narrowing. 

\subsection{Experimental Recipe\label{sub:experimentalrecipe}}

\begin{figure}[h!]
\scalebox{0.45}{\includegraphics{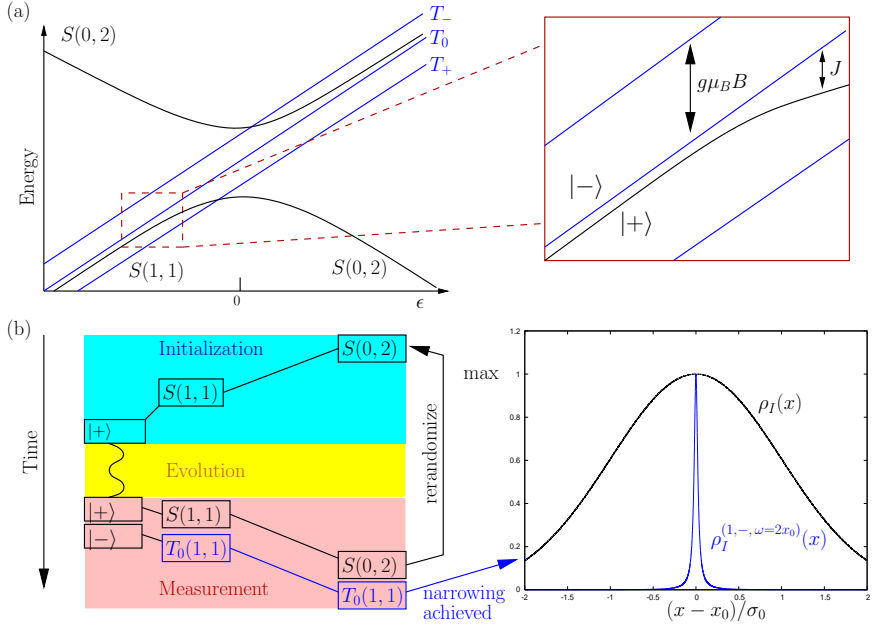}}
\caption{In this figure the pulse-sequence for one measurement in the 
basis $\ket{\pm}$ is explained. (a) The level diagram for the
two-electron spin states (sweeping $\epsilon=E_{S(1,1)}-E_{S(0,2)}$
with $E_{S(1,1)}+E_{S(0,2)}$ held constant). 
Inset: The splitting between the $S(1,1)$ and $T_0$
state is given by the exchange interaction $J$ between the two
dots. For $J \rightarrow 0$ the $\ket{\pm}$ states become
eigenstates. (b) The change of the detuning
$\epsilon$ during the course of 
the measurement is sketched: the position of the boxes corresponds to
the value of $\epsilon$. After applying the oscillating signal the
system is in either one of $\ket{\pm}$, which results in a different
state when switching back to positive detuning. If the system ends up in the $T_0$ state
(which corresponds to measurement result $\ket{-}$) narrowing has been
achieved, otherwise the nuclear system must be rerandomized and
the measurement repeated. \label{fig:pulsesequence}}
\end{figure}

An experimental implementation of this scheme of course requires the
ability to initialize to the state $\ket{+}$ and to read-out the states
$\ket{\pm}$. This has recently been achieved in an experiment by Petta
 \emph{et al.} \cite{Petta:2005b} using adiabatic passage from the
$S^z=0$ singlet. What needs to be achieved in addition is to apply an
external magnetic field gradient $\delta b^z$ between the two dots in order to
satisfy the requirements on the parameters of the system:
\begin{equation}\label{eq:requirements}
J_0\ll x_0,\,\,\, j\ll x_0,\,\,\, \sigma_0\ll x_0,\,\,\, j<\sigma_0.
\end{equation}
Typical values for the parameters satisfying these requirements are:
$1/\sigma_0=10$ ns, $1/j=100$ ns,
$\omega=2x_0=10^9-10^{10}$ Hz. 

The pulse-sequence for one measurement is shown in
Fig. \ref{fig:pulsesequence}. The parameter $\epsilon$ describes the
detuning between the singlet state with two electrons on the right
dot and the singlet state with one electron on each dot:
$\epsilon=E_{S(0,2)}-E_{S(1,1)}$. First the system is set to the $S(1,1)$ from
the $S(0,2)$ state by going from large positive to negative detuning
$\epsilon$ (such that still $J\gg |x_0|$) as described in
Ref. \cite{Petta:2005b} (rapid adiabatic passage through $S(1,1)-T_+$
resonance) . In the limit of 
$J \ll |x_0|$ and $x_0>0$, the ground state is $\ket{+}$ (for $x_0<0$,
the ground state is $\ket{-}$ and $\ket{\pm}$ thus need to be interchanged
in the following description) and 
initialization to $\ket{+}$ is thus possible by adiabatic passage from
$S(1,1)$, i.e., by switching adiabatically to large negative
detuning (such that $J\ll x_0$). Then the oscillating signal is applied
to $J(t)$ for a time $t_m$. Finally we adiabatically switch back to 
$J\gg x_0$. With this the $\ket{+}$ state goes to the singlet $S(1,1)$, and the
$\ket{-}$ state goes to the $S^z=0$ triplet $T_0(1,1)$. Read-out of
the singlet and triplet may then be achieved via switching to large
positive detuning: the $S(1,1)$ state goes over to the $S(0,2)$ while
the $T_0(1,1)$ does not since tunneling preserves spin. The number of electrons
on the right dot can then be detected via a charge sensor (QPC). 
If the outcome of the measurement is $\ket{-}$, we have achieved
narrowing. In the case where we have measured $\ket{+}$, the nuclear
spin distribution is rerandomized and the measurement is repeated. 

\subsection{Adaptive scheme\label{sub:adaptivescheme}}
If measurements at several different driving frequencies can be
performed, a systematic narrowing of the distribution can be achieved
by an adaptive scheme. Such an adaptive scheme is more intricate than
the one described above, but allows one to narrow by more than a
factor $j/4\sigma_0$. 

\begin{figure}[h!]
\scalebox{0.58}{\includegraphics{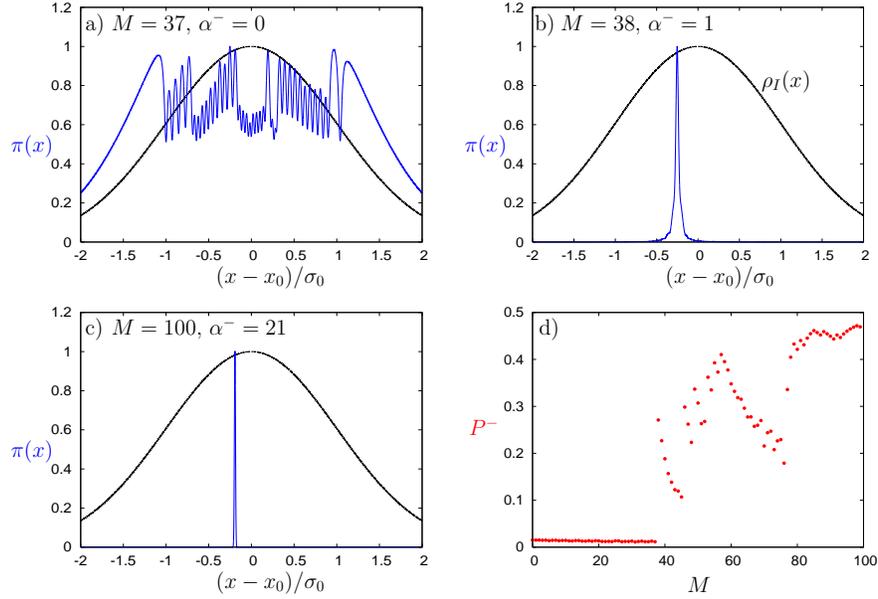}}
\caption{\label{fig:adaptive}In this figure we show a typical \cite{typical} sequence of
  the rescaled probability density of eigenvalues
  $\pi(x)=\rho_I^{(M,\{\alpha^-_i\},\{\omega_i\})}(x)/
  \mathrm{max}\left(\rho_I^{(M,\{\alpha^-_i\},\{\omega_i\})}(x)\right)$
  for the adaptive scheme.        
  Here, $\rho_I^{(M,\{\alpha^-_i\},\{\omega_i\})}(x)$ is given
  in Eq.(\ref{Eq:rhomultifreq}). We have 
  $x=\delta h^z_n+\delta b^z$, $j/\sigma_0=1/10$,
  $\alpha^-=\sum_{i=1}^M \alpha^-_i$, and in a)--c) the 
  initial Gaussian distribution (with FWHM $2\sigma_0\sqrt{2\ln
    2}\approx 2\sigma_0$) is plotted for reference.
 a) Up to $M=37$ measurements
the outcome is never $\ket{-}$ and thus each measurement ``burns a
hole'' into the distribution where it previously had its maximum. b)
In the $38th$ measurement the outcome is $\ket{-}$
which leads to a narrowed distribution of nuclear spin eigenvalues
(peak centered at $\approx -0.25$) with a FWHM that is reduced by a factor
$\approx j/4 \sigma_0$. c) Adapting the 
driving frequency $\omega$ to this peak, i.e., setting $\omega/2=x_{\mathrm{max}}$ in subsequent
measurements, leads to further narrowing
every time $\ket{-}$ is measured. In
this example the final FWHM is $\approx\sigma_0/100$, i.e., the
distribution has been narrowed by a 
factor $\approx j/10\sigma_0$. d) The probability $P^-$ to 
measure $\ket{-}$ jumps up after the $38th$ measurement and after
$\ket{-}$ is measured several more times, this probability saturates close to $1/2$.} 
\end{figure}

The results of Eqs. (\ref{Eq:rhoplus},\ref{Eq:rhominus}) may be generalized
to the case of $M$ subsequent measurements at different driving
frequencies $\omega_i$ under the assumption, that the nuclear spin
system is static between subsequent measurements:
\begin{equation}\label{Eq:rhomultifreq}
\rho_I^{(M,\{\alpha^-_i\},\{\omega_i\})}(x)=\frac{\rho_I(x)}
  {Q(\{\alpha^-_i\},\{\omega_i\})}\prod^M_{i=1}
  L_{\omega_i}^{\alpha^-_i}(1-L_{\omega_i})^{1-\alpha^-_i},
\end{equation}
where $Q(\{\alpha^-_i\},\{\omega_i\})$ is the normalization factor, 
$\alpha^-_i=1$ for measurement outcome $\ket{-}$ and $\alpha^-_i=0$
for measurement outcome $\ket{+}$ in the $\mathrm{i^{th}}$ measurement
with driving frequency $\omega_i$. Further, $\{\omega_i\} = \{\omega_1,
\dots ,\omega_{M}\}$ and $\{\alpha^{-}_i\}=\{\alpha^{-}_1, \dots ,
\alpha^{-}_{M}\}$. As we have seen in the case of just one
measurement, it is the measurement outcome $\ket{-}$ that leads to
narrowing. Thus, before each measurement $\omega_i$ is chosen 
to maximize the probability $P^-_{\omega_i}$ to measure $\ket{-}$. The
reason that $\omega_i$ must be adapted and that one should not keep
measuring at the same driving frequency is that the measurement
outcome $\ket{+}$ causes a dip in $\rho_I(x)$ at the position where $L_{\omega_i}(x)$
has its peak and since $P^-_{\omega_i}$ is the overlap of $\rho_I(x)$
and  $L_{\omega_i}(x)$, this causes  $P^-_{\omega_i}$ to diminish with each measurement. 

To see what is a typical measurement history for such an adaptive
scheme we have performed simulations. The results for a typical \cite{typical}
sequence of measurements is shown in Fig. \ref{fig:adaptive} (for
another sequence see Fig.2 in Ref. \cite{Klauser:2006a}).

\section{Conclusions\label{sec:conclusions}}

We have reviewed our scheme that uses pseudospin measurements in the 
$S^z=0$ subspace of two electron spin states in a double quantum dot to narrow
the distribution of difference in nuclear polarization between the two dots. 
A successful experimental implementation of this scheme would allow to suppress 
hyperfine-induced decoherence and thus to reach the coherence times required for 
efficient quantum error correction.

\section*{Acknowledgements}
We thank G. Burkard, M. Duckheim, J. Lehmann, F. H. L. Koppens,
D. Stepanenko and, in particular, A. Yacoby for useful
discussions. We acknowledge financial support from the Swiss NSF,
the NCCR nanoscience, EU NoE MAGMANet, DARPA, ARO, ONR, JST ICORP,
and NSERC of Canada.

\bibliographystyle{prsty}    
\bibliography{Klauser}      

\begin{thebibliography}{10}
\bibitem{assumption}
{This assumption is not necessary for our narrowing scheme. However, it does
  allow for the derivation of the analytical formulas in this section, which
  give insight into the mechanism of narrowing.}
\bibitem{typical}
{We have performed more than 60 runs of the simulation, varying $M$ and
  $j/\sigma_0$}.
\end{thebibliography}

\end{document}